\documentclass[twocolumn,showpacs,amsmath,amssymb]{revtex4}
\usepackage{mathrsfs}
\usepackage{graphicx,color}
\usepackage{dcolumn}
\usepackage{bm}
\usepackage{CJK}
\usepackage{epstopdf}

\begin{document}
\begin{CJK*}{GBK}{song}
\title{Construction of Nuclear Covariant Energy Density Functional from A Physics-Guaranteed Neural Network Approach}

\author{W. F. Li$^{1}$}
\author{Z. M. Niu$^{1}$}\email{zmniu@ahu.edu.cn}
\author{H. Z. Liang$^{2,3,4}$}
\author{Y. F. Niu$^{5}$}
\author{B. H. Sun$^{6}$}

\affiliation{$^1$School of Physics, Anhui University,
             Hefei 230601, China}
\affiliation{$^2$Department of Physics, Graduate School of Science, The University of Tokyo,
             Tokyo 113-0033, Japan}
\affiliation{$^3$Quark Nuclear Science Institute, The University of Tokyo, Tokyo 113-0033, Japan}
\affiliation{$^4$RIKEN Center for Interdisciplinary Theoretical and Mathematical Sciences (iTHEMS), Wako 351-0198, Japan}
\affiliation{$^5$School of Physics and Astronomy, Shanghai Jiao Tong University, Key Laboratory for Particle Astrophysics and Cosmology (MoE), Shanghai Key Laboratory for Particle Physics and Cosmology, Shanghai 200240, China}
\affiliation{$^6$School of Physics and Nuclear Energy Engineering, Beihang University,
             Beijing 100191, China}

\begin{abstract}
Density functional theory is a practical approach for solving quantum many-body problems with available computational resources. The complexity of the nuclear force makes constructing an accurate nuclear energy density functional much more challenging. The feasibility of constructing a nuclear covariant energy density functional with deep neural networks is demonstrated. This physics-guaranteed neural network approach achieves high accuracy in predicting nuclear energy density and exhibits significantly better extrapolation abilities than traditional machine learning methods for binding energies. When combined with the existing covariant density functional, the neural network approach improves the binding energy accuracy from $644$ keV to $86$ keV in the known region and also effectively captures the microscopic shell effect. Furthermore, its extrapolation performance is also significantly enhanced, achieving an accuracy of approximately $5$ MeV even when extrapolating up to $30$ steps. This work paves the way for the construction of accurate nuclear energy density functionals through machine learning.
\end{abstract}

\maketitle

\section{Introduction}
The exact solution of quantum many-body problems is a significant challenge in various fields of physics~\cite{March1967ManyBody}, including nuclear physics, condensed matter physics, and quantum chemistry. Density functional theory (DFT) provides a practical approach for solving this problem with available computational resources~\cite{Dreizler1990DFT, Kohn1999RMP}, with great success in areas such as materials science~\cite{Souza2007NL, Wisesa2025NL}, quantum chemistry~\cite{Mardirossian2017MP}, and condensed matter physics~\cite{Tao2016PRL}. The Hohenberg-Kohn theorem guarantees the existence of such a functional~\cite{Hohenberg1964PR}. In practice, the formulation is commonly implemented through the Kohn-Sham equations~\cite{Kohn1965PR}, wherein an interacting many-body system is mapped onto a non-interacting reference system, which provides the same ground-state density with a local Kohn-Sham potential. In nuclear physics, the complexity of the nuclear force makes it much more challenging to construct an accurate nuclear energy density functional. The accuracy of existing energy density functionals is generally greater than $1$ MeV in the known region, whereas the deviations among them even reach tens of MeV when extrapolated to the region near the neutron-drip line~\cite{Bender2003RMP}. Consequently, the development of an accurate nuclear energy density functional has become one of the central topics in nuclear theory.

Nuclear covariant density functional theory (CDFT) originates from relativistic quantum field theory, in which a system of Dirac nucleons interacts in a relativistic covariant manner via the meson and the photon fields~\cite{Meng2006PPNP, Meng2016RDFT}. Compared with the non-relativistic DFT, the CDFT includes not only inherent Lorentz covariance, but also the automatic incorporation of the nucleon spin degrees of freedom, a natural explanation for the pseudospin symmetry in nucleon spectrum~\cite{Ginocchio1997PRL, Meng1998PRC, Meng1999PRC, Chen2003CPL, Ginocchio2005PR, Liang2015PR} and the spin symmetry in antinucleon spectrum~\cite{Liang2015PR, Zhou2003PRL, He2006EPJA}, as well as the inherent inclusion of nuclear magnetic potential~\cite{Koepf1989NPA}, among others. The CDFT has attracted much attention and successfully describes many nuclear properties~\cite{Ring2012PS}, establishing itself as one of the most prominent microscopic frameworks in nuclear theory~\cite{Ring1996PPNP, Vretenar2005PR, Niksic2011PPNP, Meng2013FP, Meng2015JPG, Zhou2016PS, Shen2019PPNP}.

In recent years, machine learning (ML) methods have provided powerful tools in physics, including particle physics~\cite{Baldi2014NC, Pang2018NC, Brehmer2018PRL}, condensed matter physics~\cite{Carrasquilla2017NP, Carleo2017Science}, and astrophysics~\cite{Navarro2021ApJ, Navarro2022ApJS}. The ML methods have also been used to predict various nuclear properties, often achieving higher predictive accuracy than those obtained from conventional theoretical models~\cite{Bedaque2021EPJA, Boehnlein2022RMP, He2023SCPMA}. The ML methods have also demonstrated significant potential in enhancing the accuracy of DFT functionals~\cite{Pederson2022NRP}. In electronic structure calculations, the ML has been successfully employed to develop functional forms with high accuracy and transferability, even addressing challenges that are difficult for conventional approaches, such as fitting molecular dissociation curves and describing strongly correlated systems~\cite{Brockherde2017NC, Li2021PRL}. The DM21 functional, developed by DeepMind, utilizes a neural network architecture trained on large-scale datasets and achieves performance superior to most human-designed functionals in molecular systems, while also exhibiting strong generalization abilities~\cite{Kirkpatrick2021Science}. These advancements demonstrate that ML can not only automatically extract complex features from data but also effectively incorporate physical constraints, thereby driving innovation in functional design~\cite{Li2021PRL, Nagai2020npjCM}. The complexity of nuclear force makes it much more difficult to construct an accurate DFT for self-bound nuclear systems. In this work, we attempt to construct a nuclear energy density functional directly using ML methods. Such ML models guaranteed by physics theory may be promising to achieve higher accuracy and better extrapolation abilities.

The CDFT can be constructed with the relativistic mean-field (RMF) model in the meson-exchange or point-coupling representations. The point-coupling RMF model provides more explicit expressions of energy density in terms of nucleon densities, while the relativistic covariant advantages in CDFT still remain~\cite{Nikolaus1992PRC, Burvenich2002PRC, Zhao2010PRC, Zhao2022PRC, Liu2023PLB}. In this letter, a successful attempt for constructing a nuclear covariant energy density functional in the point-coupling representation is made with deep neural networks. This framework establishes a novel paradigm for ML applications in nuclear physics and is promising to achieve better prediction and generalisation abilities. In the following, the basic formulas of CDFT and the architecture of the ML method will be briefly introduced. We will then investigate the feasibility, as well as the prediction and generalization abilities of this physics-guaranteed ML method by comparing it with traditional ML methods.

\section{Theoretical framework}
The starting point of the relativistic point-coupling model is the effective Lagrangian density
\begin{eqnarray}\label{Eq:Lagrangian}
{\cal L} = {\cal L}_{\rm free}+{\cal L}_{\rm 4f}+{\cal L}_{\rm hot}+{\cal L}_{\rm der}+{\cal L}_{\rm em},
\end{eqnarray}
where ${\cal L}_{\rm free}$, ${\cal L}_{\rm 4f}$, ${\cal L}_{\rm hot}$, ${\cal L}_{\rm der}$, and ${\cal L}_{\rm em}$ represent the Lagrangian densities of the free nucleons, the four-fermion point-coupling terms, the higher-order terms, the derivative terms, and the electromagnetic interaction terms, respectively. Taking the mean-field and ``no-sea'' approximations, one can get the expectation value $\langle\Phi|H|\Phi\rangle$ of Hamiltonian $H$ corresponding to the Lagrangian density~\cite{Nikolaus1992PRC, Burvenich2002PRC, Zhao2010PRC}. Minimizing $\langle\Phi|H|\Phi\rangle$ with respect to the wave functions, the ground-state binding energy of a nuclear system can be obtained $B_{\rm CDFT} = -E_{\rm CDFT}$, where
\begin{equation}
\begin{split}\label{Eq:er}
E_{\rm CDFT} &= \int d^3r {\cal E}(\emph{\textbf r}) \\
             &= \int d^3r \left\{{\cal E}_{\rm kin}(\emph{\textbf r}) + {\cal E}_{\rm int}(\emph{\textbf r}) + {\cal E}_{\rm em}(\emph{\textbf r})\right\}.
\end{split}
\end{equation}
${\cal E}_{\rm kin}$, ${\cal E}_{\rm int}$, and ${\cal E}_{\rm em}$ represent the kinetic, interaction, and electromagnetic parts of the total energy density ${\cal E}$, respectively. The even-even nuclei with $8\leqslant Z \leqslant 100$ and $8\leqslant N \leqslant 156$ are studied in this work, for which the space-like components of the currents vanish due to the time-reversal invariance. The remaining isoscalar-scalar density $\rho_S$, isoscalar-vector density $\rho_V$, isovector-vector density $\rho_{TV}$ are used to describe the energy densities in Eq.~(\ref{Eq:er}). The CDFT with PC-PK1 parameter set provides an accurate and reliable description of the isospin dependence of nuclear binding energy~\cite{Zhao2010PRC}, which has been widely used for studying nuclear properties~\cite{Zhao2012PRC, Zhang2014FP, Lu2015PRC, Zhang2021PRC}. The deep neural network is employed to construct a nuclear CDFT by taking the PC-PK1 results as the training data, which means that the CDFT with the PC-PK1 is taken as the target DFT in this work, so the deep neural network constructed in this work could serve as a DFT emulator of PC-PK1. The spherical approximation is used in the calculations of the CDFT with PC-PK1, so the densities can then be simplified as functions of the radial coordinate $r$, which is discretized with $r_i = 0.1 \times i~{\rm fm}~(i=0,1,...,150)$. The details can be seen in Supplemental Material~\cite{SuppleMater}.

Three classes of neural networks with different input layers are constructed to describe nuclear energy ${E}_{\rm CDFT}$ and energy density ${\cal E}_{\rm CDFT}$: The first class (NN1) employs the traditional neural network architecture, i.e., the neural network with the inputs $I = (N - Z) / (N + Z)$ and $P = \nu_p\nu_n/(\nu_p+\nu_n)$ besides $Z$ and $N$, where $\nu_p$ and $\nu_n$ are the differences between the nucleon numbers $Z$ and $N$ and the nearest magic numbers ($8$, $20$, $28$, $50$, $82$ for protons and $8$, $20$, $28$, $50$, $82$, $126$, $184$ for neutrons) \cite{Niu2018PLB}. The inputs of the neural networks in the second class (NN2) are various densities at all mesh points. The third class (NN3) takes various densities at the coordinate $r$ as inputs. A composite loss function $L_{\rm total}$ is designed for the networks with the output of energy density ${\cal E}_{\rm CDFT}$, which is
\begin{eqnarray}
L_{\rm total} = a L_E + L_{\cal E},
\end{eqnarray}
with,
\begin{align}
L_E &= \frac{1}{n}\sum_{j=1}^n(E_{{\rm CDFT},j}-E_{{\rm NN},j})^2,\\
L_{\cal E} &= \frac{1}{151n}\sum_{j=1}^n\sum_{i=0}^{150}[ ({\cal E}_{{\rm CDFT},j}(r_i)-{\cal E}_{{\rm NN},j}(r_i))\cdot r_i^2]^2,
\end{align}
where $a$ is the weight factor, $L_E$ and $L_{\cal E}$ are employed to assess the loss of the binding energy and energy density with respect to training data, and $n$ is the number of nuclei in the learning set. A reasonable parameter $a$ ensures that the neural network can well describe binding energy and energy density simultaneously, which is automatically adjusted by balancing $L_E$ and $L_{\cal E}$. For the neural networks with the output of energy $E$, the loss function is $L_E$. The input variables, output variables, and loss functions of various neural networks used in this work are given in Table~\ref{network}. The neural networks with the output of ${\cal E}$, i.e., NN2-I3${\cal E}$ , NN2-I6${\cal E}$, NN3-I3${\cal E}$, and NN3-I6${\cal E}$ construct nuclear covariant energy density functionals. Comparing with the NN2 using the global densities as input features, the weight-sharing architecture of NN3 guarantees that the neural network parameters are the same for the densities at different spatial coordinates. Furthermore, the current neural network architecture is implemented using a fully differentiable framework. By utilizing automatic differentiation, the functional derivative of the total energy with respect to the nucleon densities can be obtained efficiently. This differentiability ensures that the proposed framework is theoretically feasible with the variational principle in density functional theory and can be directly integrated into future self-consistent calculations. Therefore, the physics-guaranteed covariant density functional is realized by NN3 through its architectural design and the underlying differentiable programming. The Adaptive Moment Estimation (Adam)~\cite{Adam2014arXiv} optimizer from PyTorch~\cite{PyTorch} is used for parameter optimization. The network framework and the training process for these neural networks are described in detail in Supplemental Material~\cite{SuppleMater}.

\begin{table*}
\centering
\caption{Input variables, output variables, and loss functions of various neural networks. ``size:1*151'' or ``size:1*1'' indicates that the input/output variable is density at all mesh points or density at the coordinate $r$. All the densities ($\rho_S$, $\rho_V$, $\rho_{TV}$, $\Delta\rho_S$, $\Delta\rho_V$, $\Delta\rho_{TV}$, and ${\cal E}$, where $\Delta$ is the Laplace operator) involved are multiplied by $r^2$ to improve the performance of neural networks. The results of $\rho_x$, $\Delta\rho_x$ ($x = S, V, TV$) are calculated by the CDFT with PC-PK1~\cite{Zhao2010PRC}.}
\begin{tabular}{cccc}
\hline
~Neural network~ & ~Input variables~ & ~Output variables~ & ~Loss function~ \\
\hline
NN1-I4E                 & $Z$, $N$, $I$, $P$ (size:4*1)                                                                   & $E$ (size:1*1)          & $L_E$ \\
NN2-I3E                 & $\rho_S$, $\rho_V$, $\rho_{TV}$ (size:3*151)                                                    & $E$ (size:1*1)          & $L_E$ \\
NN2-I6E                 & $\rho_S$, $\rho_V$, $\rho_{TV}$, $\Delta\rho_S$, $\Delta\rho_V$, $\Delta\rho_{TV}$ (size:6*151) & $E$ (size:1*1)          & $L_E$ \\
NN2-I3${\cal E}$        & $\rho_S$, $\rho_V$, $\rho_{TV}$ (size:3*151)                                                    & ${\cal E}$ (size:1*151) & $L_{\rm total}$ \\
NN2-I6${\cal E}$        & $\rho_S$, $\rho_V$, $\rho_{TV}$, $\Delta\rho_S$, $\Delta\rho_V$, $\Delta\rho_{TV}$ (size:6*151) & ${\cal E}$ (size:1*151) & $L_{\rm total}$ \\
NN3-I3${\cal E}$        & $\rho_S$, $\rho_V$, $\rho_{TV}$ (size:3*1)                                                      & ${\cal E}$ (size:1*1)   & $L_{\rm total}$ \\
NN3-I6${\cal E}$        & $\rho_S$, $\rho_V$, $\rho_{TV}$, $\Delta\rho_S$, $\Delta\rho_V$, $\Delta\rho_{TV}$ (size:6*1)   & ${\cal E}$ (size:1*1)   & $L_{\rm total}$ \\
\hline
NN1-I4$\delta$E         & $Z$, $N$, $I$, $P$ (size:4*1)                                                                   & $E-E_{\rm PC-PK1}$ (size:1*1)                 & $L_E$ \\
NN2-I6$\delta$E         & $\rho_S$, $\rho_V$, $\rho_{TV}$, $\Delta\rho_S$, $\Delta\rho_V$, $\Delta\rho_{TV}$ (size:6*151) & $E-E_{\rm PC-PK1}$ (size:1*1)                 & $L_E$ \\
NN2-I6$\delta {\cal E}$ & $\rho_S$, $\rho_V$, $\rho_{TV}$, $\Delta\rho_S$, $\Delta\rho_V$, $\Delta\rho_{TV}$ (size:6*151) & ${\cal E}-{\cal E}_{\rm PC-PK1}$ (size:1*151) & $L_{\rm total}$ \\
NN3-I6$\delta {\cal E}$ & $\rho_S$, $\rho_V$, $\rho_{TV}$, $\Delta\rho_S$, $\Delta\rho_V$, $\Delta\rho_{TV}$ (size:6*1)   & ${\cal E}-{\cal E}_{\rm PC-PK1}$ (size:1*1)   & $L_{\rm total}$ \\
\hline
\end{tabular}
\label{network}
\end{table*}

The nuclei in the known region, i.e., the nuclei with the experimental masses in AME2020~\cite{Wang2021CPC}, are randomly divided into $20$ different groups, each containing a learning set ($80\%$ of the data) and a validation set ($20\%$ of the data) to reduce potential bias arising from a single randomization. This repeated sampling approach ensures statistical robustness in model development and performance evaluation. Each of these $20$ groups is trained using five random initial parameters. Neural networks are employed to predict nuclear energy densities and binding energies, and the final predictions are derived from the statistical mean of $100$ independent realizations.

\section{Results and Discussion}
\begin{figure}
\includegraphics[width=8.6cm]{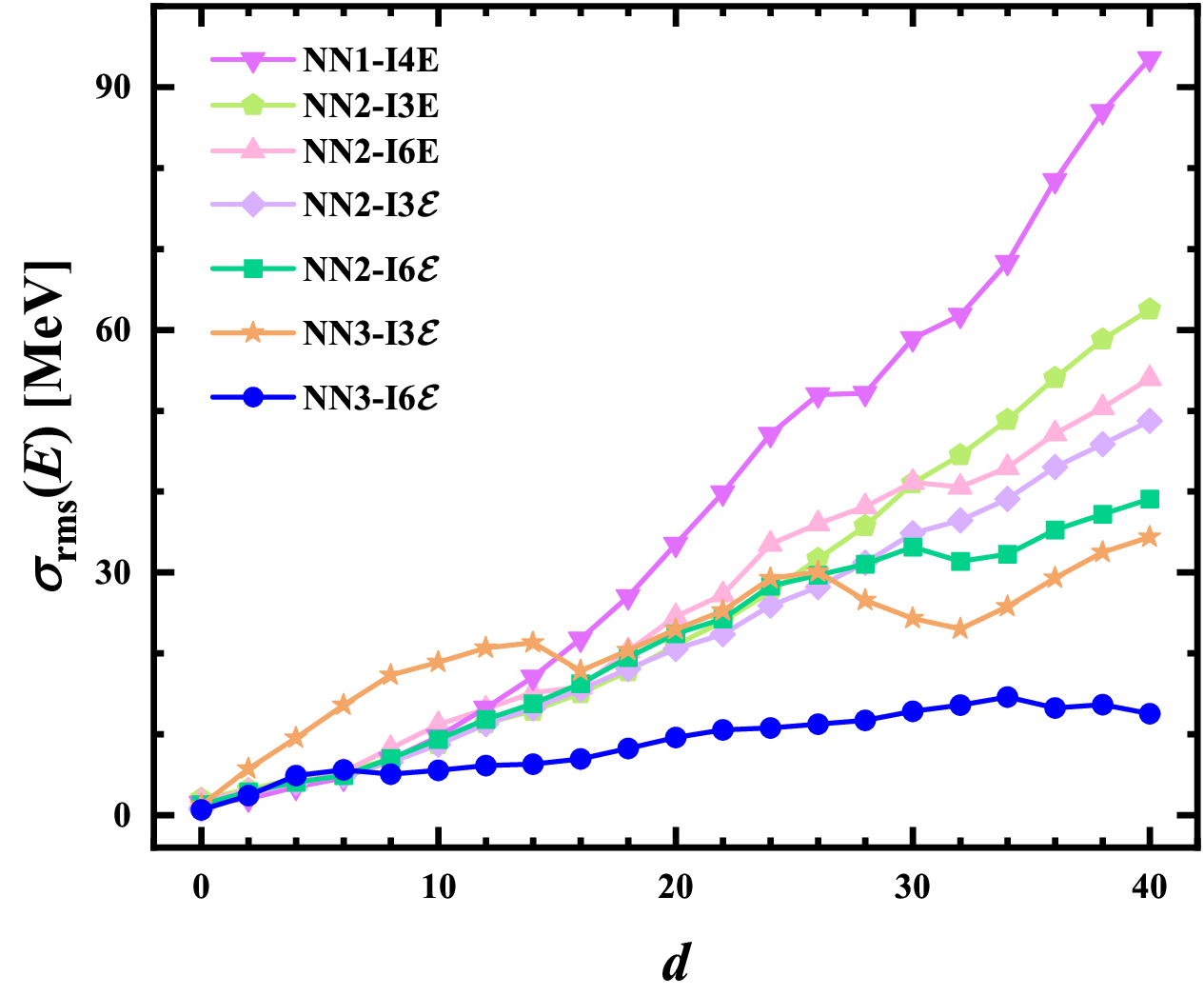}
\caption{(Color online) Rms deviations $\sigma_{\rm rms}(E)$ of the binding energies predicted by various neural networks with respect to the target data from the calculations with PC-PK1 as a function of the minimum distance $d$ to the isotopes in the known region.}
\label{Fig:SigmaR1}
\end{figure}
To investigate the extrapolation abilities of different neural networks, the root-mean-square (rms) deviations $\sigma_{\rm rms}(E)$ of the predicted binding energies with respect to the target data from the PC-PK1 calculations are shown in Fig.~\ref{Fig:SigmaR1} as a function of the minimum distance $d$ to the isotopes in the known region. $N = 156$ is the maximum neutron number in all learning sets, so only the nuclei with $N \leqslant 156$ are included in our studies. Clearly, $\sigma_{\rm rms}(E)$ generally increases as the $d$ increases for all networks. Compared to NN1, which is widely used in nuclear mass predictions, NN2 and NN3 exhibit much better extrapolation abilities, which indicates that more physics can be extracted by the neural network from various densities. Notably, NN3-I6${\cal E}$ employing point-to-point mapping achieves the both optimal prediction accuracy with the $\sigma_{\rm rms}(E)$ of $644$~keV in the known region and best extrapolation ability by constructing an energy density functional, i.e., the density $\rho_{S,V,TV}$ and $\Delta\rho_{S,V,TV}$ at different $r$ uses the same neural network to get the energy density ${\cal E}(r)$. The best extrapolation ability implies that this neural network architecture can better capture the essential physics correlations in the description of nuclear properties. In principle, this approach could also be used to directly learn the experimental binding energies and then generate a new functional form. However, this requires extending the approach to incorporate deformation degrees of freedom and further introduce various corrections beyond the mean-field approximation, such as vibrational and rotational corrections. Therefore, constructing a new nuclear energy density functional by learning the experimental binding energies remains a significant challenge at present. The densities $\rho_{S,V,TV}$ are clearly not completely independent of each other, while it is intractable in defining the specific relationship among them. Once the relationship among the densities $\rho_{S,V,TV}$ is established, the energy density functional constructed in this work can obtain ground-state observables through the variational of density, such as charge radii and density distribution. Due to the complex relationship among the densities $\rho_{S,V,TV}$, this work calculates the ground-state energies based on the constructed functional and the known ground-state densities instead of solving the ground-state densities via variational methods.

\begin{figure}
\includegraphics[width=8.6cm]{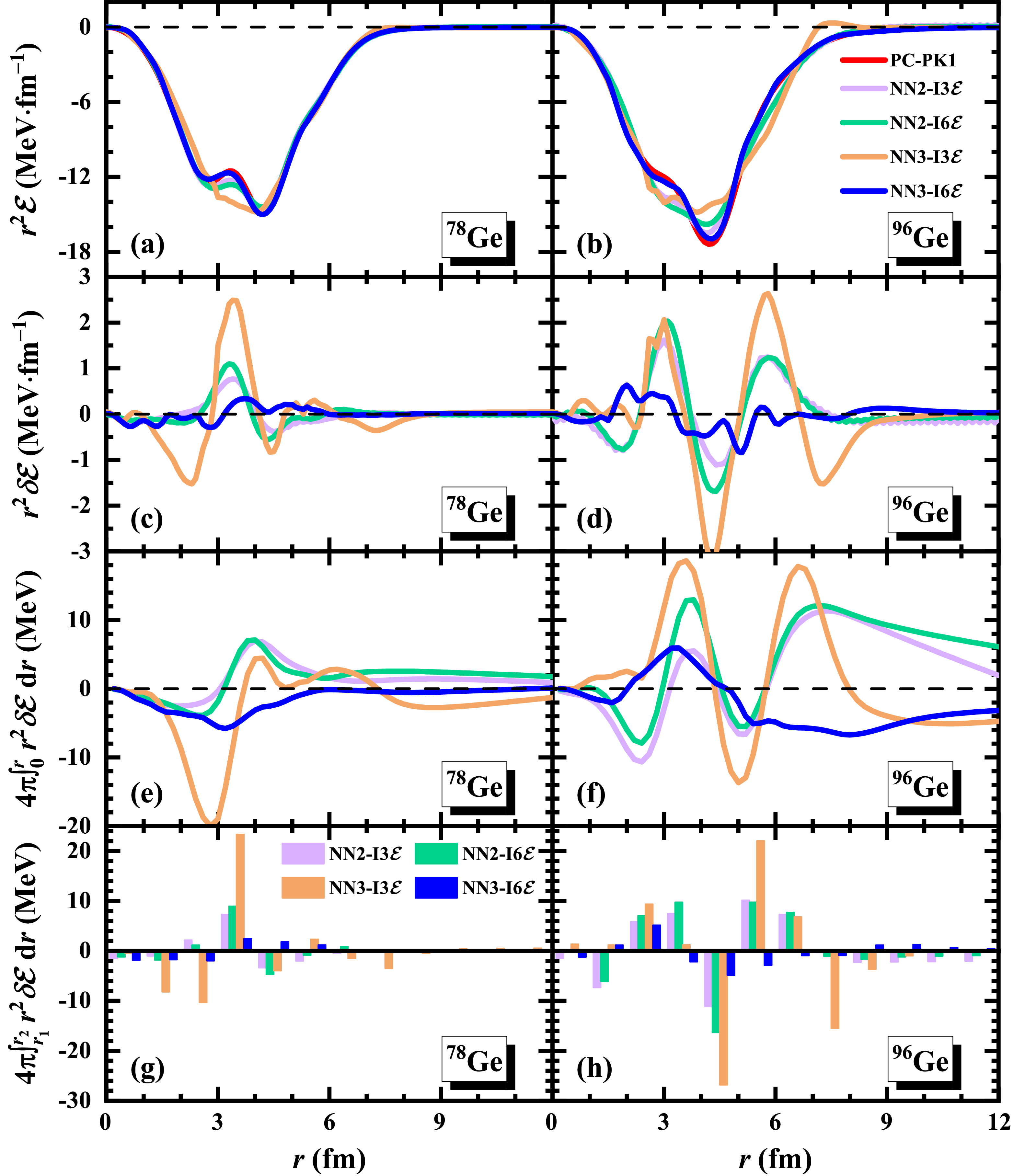}
\caption{(Color online) Benchmarking of the nuclear energy density predicted by various neural networks against the target PC-PK1 calculations for $^{78,96}$Ge. Panels (a,b), (c,d), and (e,f) correspond to the $r^2$-weighted energy densities $r^2{\cal E}$, the differences of $r^2{\cal E}$ between PC-PK1 calculations and the neural network predictions, i.e., $r^2 \delta{\cal E}$, and the cumulative integral of $r^2\delta{\cal E}$ from $0$ to $r$, respectively. The histograms located between coordinates $r_1$ and $r_2$ in the panels (g, h) represent the subsection integral of $r^2\delta{\cal E}$ from $r_1$ to $r_2$.}
\label{Fig:Er}
\end{figure}
Taking $^{78}$Ge (a nucleus in the known region) and $^{96}$Ge (a nucleus extrapolated by $10$ steps beyond the known region) as examples, Fig.~\ref{Fig:Er} compares the predictions of energy densities based on four different neural networks. It is clear that NN3-I6${\cal E}$ best reproduces the target energy densities for both $^{78}$Ge and $^{96}$Ge, while the other neural networks show remarkable deviations from the PC-PK1 energy densities, particularly near the extrema and inflection points of energy densities. To highlight these behaviors, panels (c) and (d) show the differences $r^2\delta {\cal E}$ $(\delta{\cal E} = {\cal E}_{\rm PC-PK1} - {\cal E}_{\rm NN})$ between the PC-PK1 calculations and the neural network predictions. There are positive and negative $r^2\delta {\cal E}$ in different $r$, so the differences between the binding energies predicted by the PC-PK1 and the neural networks can be reduced due to cancellations. The cumulative integrals of the energy density differences $E(r)=4\pi\int_0^r r^2\delta {\cal E} dr$ are shown in panels (e) and (f). For $^{78}$Ge, $E(r)$ generally first decreases to a minimum value within $-6$ MeV, then increases, and gradually stabilizes after $r \gtrsim 6$ fm. However, the minimum value of $E(r)$ for NN3-I3${\cal E}$ is approximately $-20$ MeV, and it stabilizes only when $r \gtrsim 8$ fm. For $^{96}$Ge, $E(r)$ fluctuates considerably over the entire $r$ space. Even when $r \gtrsim 9$ fm, where the predicted energy density ${\cal E}$ closely matches that of PC-PK1, the contribution of $r^2\delta {\cal E}$ to the total energy deviation still non-negligible partially due to the amplification of the factor $r^2$. Therefore, all the densities involved in the neural networks are multiplied by $r^2$ to improve the performance of the neural networks in this work. This also underscores the importance of accurately describing the tail of the energy density in predicting nuclear binding energies. Panels (g) and (h) show the subsection integral of $r^2\delta{\cal E}$ from $r_1$ to $r_2$, i.e., $4\pi\int_{r_1}^{r_2} r^2\delta {\cal E} dr$. There are generally larger contribution of $r^2\delta {\cal E}$ to the total energy deviation at where the $r^2$-weighted energy density $r^2{\cal E}$ is larger. However, these contributions tend to largely cancel each other out, resulting in a relatively small net deviation.
The NN3-I6${\cal E}$ consistently exhibits small deviations across all intervals, both in known and unknown regions. This further supports the feasibility of using neural networks to construct nuclear density functional, demonstrating its significant advantage in describing both nuclear energy densities and binding energies.

\begin{figure}
\includegraphics[width=8.6cm]{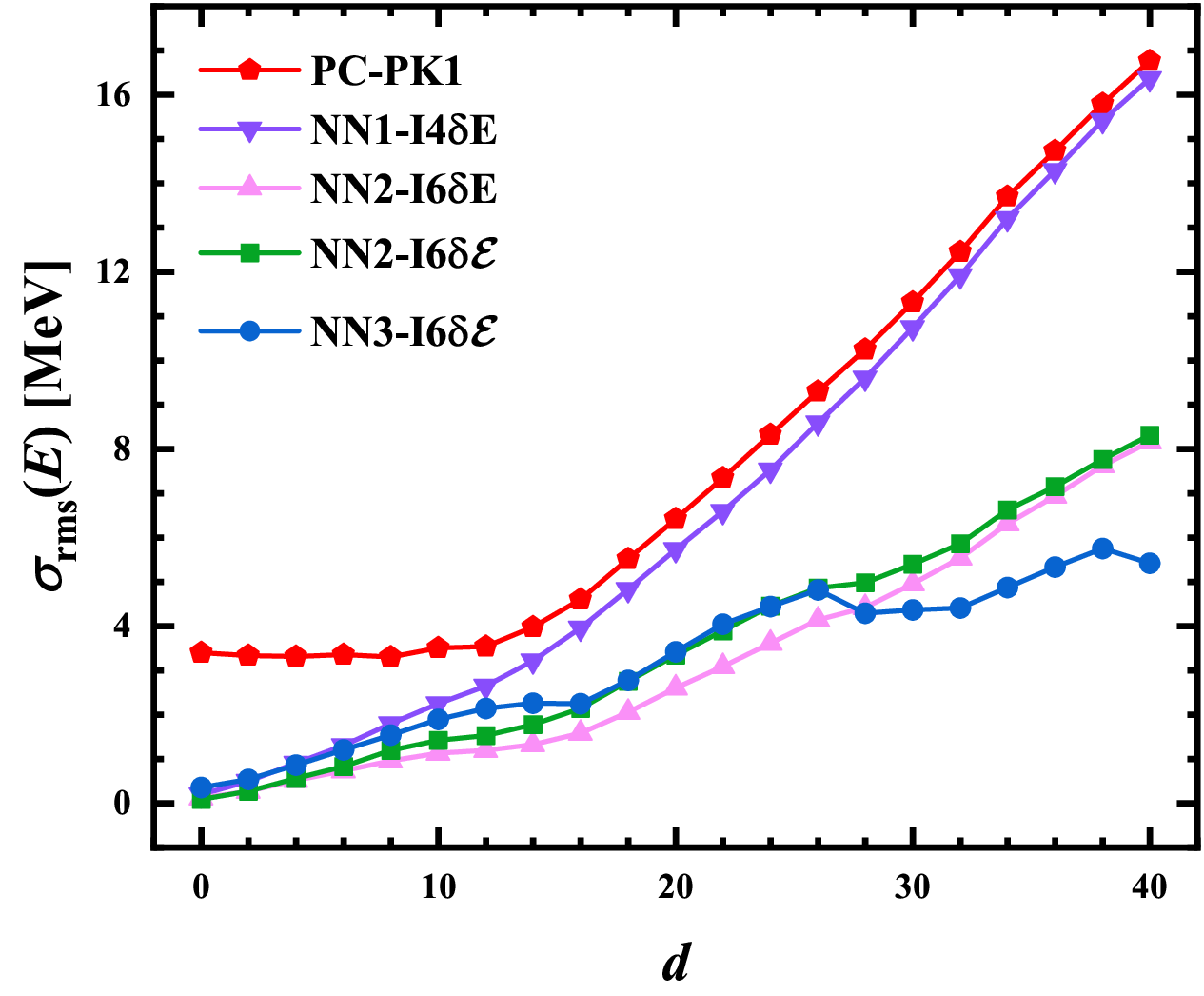}
\caption{(Color online) Same as Fig.~\ref{Fig:SigmaR1}, but the PC-F1 binding energies are taken as the target data.}
\label{Fig:SigmaR2}
\end{figure}
Numerous studies have shown that machine learning methods can significantly improve the predictive accuracy of nuclear mass models by learning the differences between experimental masses and theoretical results~\cite{Utama2016PRC, Niu2018PLB, Niu2022PRC, Wu2021PLB, Guo2024PRC}. Following this idea, we take the PC-F1 density functional results as the target data~\cite{Burvenich2002PRC} and construct four neural networks to describe the differences of binding energies or the energy densities, whose input and output variables are given in the last four rows of Table~\ref{network}. Figure~\ref{Fig:SigmaR2} shows the rms deviation $\sigma_{\rm rms}(E)$ of neural network predictions with respect to the PC-F1 binding energies as a function of the minimum distance $d$ to the isotopes in the known region.  Also shown for comparison are the rms deviations $\sigma_{\rm rms}(E)$ between the PC-F1 and PC-PK1 predictions, which exceed 3 MeV even in the known region. All four neural networks can well describe the differences between PC-F1 and PC-PK1 predictions in the known region. The $\sigma_{\rm rms}(E)$ are reduced to $0.198$, $0.118$, $0.086$, and $0.353$ MeV for NN1-I4$\delta$E, NN2-I6$\delta$E, NN2-I6$\delta{\cal E}$, and NN3-I6$\delta{\cal E}$, respectively. In the unknown region, the $\sigma_{\rm rms}(E)$ of the NN1-I4$\delta$E gradually approaches that of the PC-PK1 as $d$ increases. The neural networks with density information, i.e., NN2-I6$\delta$E, NN2-I6$\delta{\cal E}$, and NN3-I6$\delta {\cal E}$ achieve better extrapolation abilities, whose rms deviations from the target data are approximately $5$ MeV even when extrapolating up to $30$ steps. Therefore, the neural networks can well describe the energy density differences between two theoretical models even extrapolating to the unknown region, which provides a foundation for learning the differences between experimental data and theoretical predictions in the future.

\begin{figure*}
\includegraphics[width=16cm]{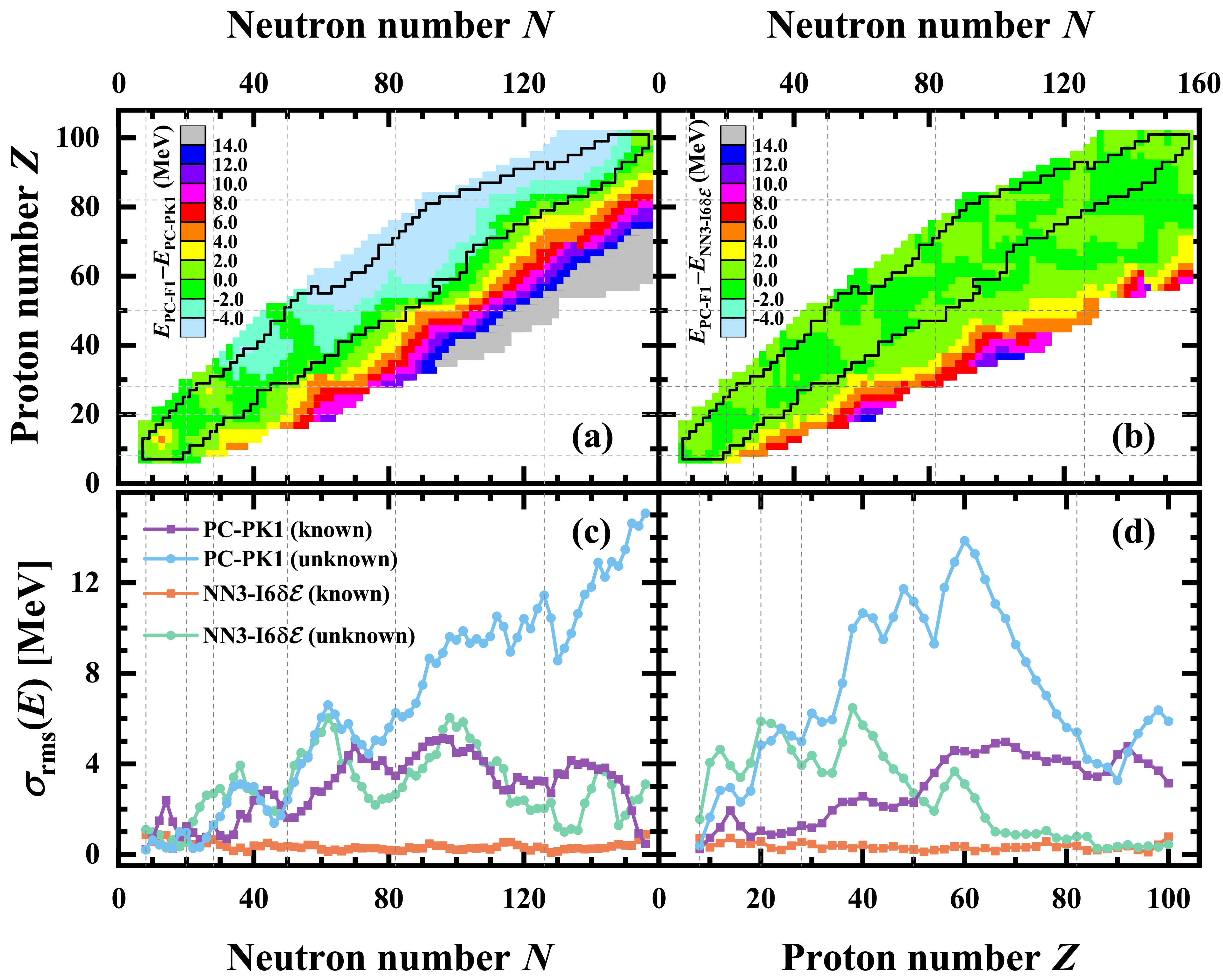}
\caption{(Color online) Comparison of the target energies $E_{\rm CDFT}$ from the PC-F1 calculations with the PC-PK1 and NN3-I6$\delta{\cal E}$ predictions. (a) Differences between the target energies $E_{\rm CDFT}$ and the PC-PK1 predictions. (b) Same as panel (a) but for the NN3-I6$\delta{\cal E}$ predictions. (c) Rms deviation $\sigma_{\rm rms}(E)$ of target energies $E_{\rm CDFT}$ with respect to the PC-PK1 and NN3-I6$\delta{\cal E}$ predictions as a function of $N$. (d) Same as panel (c) but as a function of $Z$. The solid line shows the boundary of nuclei with known masses in AME2020~\cite{Wang2021CPC} and the dashed lines denote the traditional magic numbers.}
\label{Fig:chart}
\end{figure*}
Figure~\ref{Fig:chart} shows the comparison of the target binding energies from the PC-F1 calculations with the PC-PK1 and NN3-I6$\delta{\cal E}$ predictions. As shown in Fig.~\ref{Fig:chart}(a), PC-F1 predicts larger binding energies than the PC-PK1, with the deviations exceeding $4$ MeV for the neutron-deficient nuclei with $Z \gtrsim 50$. On the neutron-rich side, the PC-F1 predicts smaller binding energies than the PC-PK1, with the deviations of larger than $20$ MeV for the nuclei near the neutron-drip line. From Fig.~\ref{Fig:chart}(b), it is clear that the NN3-I6$\delta{\cal E}$ significantly improves the description of binding energies, even for the nuclei near the drip lines. It reproduces the target binding energies generally within $1$ MeV in the known region and around $6$ MeV when extrapolating to the neutron-drip line. Figures~\ref{Fig:chart}(c) and (d) show the rms deviations of binding energies between NN3-I6$\delta{\cal E}$ predictions and target data as a function of $N$ and $Z$, respectively. Clearly, the NN3-I6$\delta{\cal E}$ significantly improves the description of binding energies for the medium and heavy mass nuclei in both the known and unknown regions, achieving comparable accuracy across the light, medium, and heavy mass regions. Surprisingly, NN3-I6$\delta{\cal E}$ shows better agreement with the target data in the unknown regions near the magic numbers. Further study finds that NN3-I6$\delta{\cal E}$ well describes the abrupt decrease in two-neutron separation energies after the magic numbers, indicating that the microscopic shell effect can be effectively captured by the NN3-I6$\delta{\cal E}$. Therefore, the neural network approach is a powerful tool for constructing a high-precision nuclear energy density functional if there are enough experimental data to train it.

The uncertainty of the neural network predictions can be assessed using the standard deviation of the $100$ independent results. It is found that the corresponding uncertainty increases as the distance from the known region increases, which is an inherent property of data-driven models. A quantitative discussion of the uncertainty in the NN3-I6${\cal E}$ and NN3-I6$\delta {\cal E}$ predictions is provided in Supplemental Material~\cite{SuppleMater}. Bayesian methods are powerful tools for model selection and uncertainty quantification. The Bayesian analysis of nuclear dynamics framework provides the conceptual scaffolding for model comparison and averaging in nuclear many-body calculations, e.g., nuclear masses and reaction cross sections~\cite{Phillips2021JPG}. Moreover, Bayesian model averaging (BMA) has been successfully used to resolve disagreements between different density functionals when extracting the nuclear symmetry energy~\cite{Qiu2024PLB}. Furthermore, BMA has been widely used to improve nuclear mass predictions and quantify extrapolation uncertainties toward the drip lines~\cite{Kejzlar2020JPG, Neufcourt2020PRC}. More recently, integrating of Bayesian model mixing with a multireference energy density functional has demonstrated significant improvements in predictive accuracy of two-neutron separation energy compared to individual theoretical models~\cite{Sharma2025PRR}. Therefore, it is promising to quantify uncertainties of our results with the Bayesian method though more computing resources are indispensable.

\section{Summary and Perspectives} The feasibility of constructing a nuclear covariant energy density functional using deep neural networks is demonstrated. By employing point-to-point mapping neural network architecture, this neural network approach guaranteed by the density functional theory achieves high accuracy in predicting both nuclear energy densities and binding energies. It also exhibits significantly superior extrapolation abilities compared to traditional neural network models. When combined with the existing covariant density functional, the binding energy accuracy is improved from $644$ keV to $86$ keV in the known regions, and the microscopic shell effect can also be effectively captured. Even when extrapolating up to $30$ steps beyond the learning region, the deviation remains about $5$ MeV, highlighting its robust generalization performance. Furthermore, this method significantly improves the description of binding energies for the medium and heavy mass nuclei, achieving comparable accuracy across the light, medium, and heavy mass regions. Although the present theoretical calculations are used as the benchmark, this approach could in principle construct a new functional form by training directly on experimental binding energies, which need the current framework be extended to incorporate deformation degrees of freedom and various beyond-mean-field corrections. Fortunately, the point-to-point mapping nature of our neural network architecture is naturally suited for generalization to deformed nuclei. This can be achieved by replacing the current one-dimensional density with two- or three-dimensional density in deformed nuclei directly, or by utilizing basis expansion coefficients of density to replace the density at coordinate space. This would establish a solid foundation for the data-driven construction of next-generation, high-precision energy density functionals in nuclear theory.

\emph{Acknowledgements.} This work was partly supported by the National Natural Science Foundation of China under Grants No. 12375109, No. 11875070 and No. 11935001, the Anhui project (Z010118169), and the Key Research Foundation of Education Ministry of Anhui Province (2023AH050095).




\end{CJK*}

\begin{thebibliography}{99}
\bibitem{March1967ManyBody} N. H. March, W. H. Young, and S. Sampanthar, The Many-Body Problem in Quantum Mechanics, Cambridge at the University Press (1967).
\bibitem{Dreizler1990DFT} R. M. Dreizler and E. K. U. Cross, Density Functional Theory: An Approach to the Quantum Many-Body Problem, Springer-Verlag (1990).
\bibitem{Kohn1999RMP} W. Kohn, Rev. Mod. Phys. \textbf{71}, 1253 (1999).
\bibitem{Souza2007NL} A. G. Souza Filho, V. Meunier, M. Terrones \emph{et al}., Nano Lett. \textbf{7}, 2383 (2007).
\bibitem{Wisesa2025NL} P. Wisesa, M. Li, M. T. Curnan, G. H. Gu, J. W. Han, J. C. Yang, and W. A. Saidi, Nano Lett. \textbf{25}, 1329 (2025).
\bibitem{Mardirossian2017MP} N. Mardirossian and M. Head-Gordon, Mol. Phys. \textbf{115}, 2315 (2017).
\bibitem{Tao2016PRL} J. Tao and Y. Mo, Phys. Rev. Lett. \textbf{117}, 073001 (2016).
\bibitem{Hohenberg1964PR} P. Hohenberg and W. Kohn, Phys. Rev. \textbf{136}, B864 (1964).
\bibitem{Kohn1965PR} W. Kohn and L. J. Sham, Phys. Rev. \textbf{140}, A1133 (1965).
\bibitem{Bender2003RMP} M. Bender, P.-H. Heenen, and P.-G. Reinhard, Rev. Mod. Phys. \textbf{75}, 121 (2003).
\bibitem{Meng2016RDFT} J. Meng, Relativistic Density Functional for Nuclear Structure, World Scientific (2016).
\bibitem{Meng2006PPNP} J. Meng, H. Toki, S. G. Zhou, S. Q. Zhang, W. H. Long, and L. S. Geng, Prog. Part. Nucl. Phys. \textbf{57}, 470 (2006).
\bibitem{Ginocchio1997PRL} J. N. Ginocchio, Phys. Rev. Lett. \textbf{78}, 436 (1997).
\bibitem{Meng1998PRC} J. Meng, K. Sugawara-Tanabe, S. Yamaji, P. Ring, and A. Arima, Phys. Rev. C \textbf{58}, R628 (1998).
\bibitem{Meng1999PRC} J. Meng, K. Sugawara-Tanabe, S. Yamaji, and A. Arima, Phys. Rev. C \textbf{59}, 154 (1999).
\bibitem{Chen2003CPL} T. S. Chen, H. F. L\"{u}, J. Meng, S. Q. Zhang, and S. G. Zhou, Chin. Phys. Lett. \textbf{20}, 358 (2003).
\bibitem{Ginocchio2005PR} J. N. Ginocchio, Phys. Rep. \textbf{414}, 165 (2005).
\bibitem{Liang2015PR} H. Z. Liang, J. Meng, and S. G. Zhou, Phys. Rep. \textbf{570}, 1 (2015).
\bibitem{Zhou2003PRL} S. G. Zhou, J. Meng, and P. Ring, Phys. Rev. Lett. \textbf{91}, 262501 (2003).
\bibitem{He2006EPJA} X. T. He, S. G. Zhou, J. Meng, E. G. Zhao, and W. Scheid, Eur. Phys. J. A \textbf{28}, 265 (2006).
\bibitem{Koepf1989NPA} W. Koepf and P. Ring, Nucl. Phys. A \textbf{493}, 61 (1989).
\bibitem{Ring2012PS} P. Ring, Phys. Scr. \textbf{T150}, 014035 (2012).
\bibitem{Ring1996PPNP} P. Ring, Prog. Part. Nucl. Phys. \textbf{37}, 193 (1996).
\bibitem{Vretenar2005PR} D. Vretenar, A. V. Afanasjev, G. A. Lalazissis, and P. Ring, Phys. Rep. \textbf{409}, 101 (2005).
\bibitem{Niksic2011PPNP} T. Nik\v{s}i\'{c}, D. Vretenar, and P. Ring, Prog. Part. Nucl. Phys. \textbf{66}, 519 (2011).
\bibitem{Meng2013FP} J. Meng, J. Peng, S. Q. Zhang, and P. W. Zhao, Front. Phys. \textbf{8}, 55 (2013).
\bibitem{Meng2015JPG} J. Meng and S. G. Zhou, J. Phys. G: Nucl. Part. Phys. \textbf{42}, 093101 (2015).
\bibitem{Zhou2016PS} S. G. Zhou, Phys. Scr. \textbf{91}, 063008 (2016).
\bibitem{Shen2019PPNP} S. H. Shen, H. Z. Liang, W. H. Long, J. Meng, and P. Ring, Prog. Part. Nucl. Phys. \textbf{109}, 103713 (2019).
\bibitem{Baldi2014NC}P. Baldi, P. Sadowski, and D. Whiteson, Nat. Commun. \textbf{5}, 4308 (2014).
\bibitem{Pang2018NC}L. G. Pang, K. Zhou, N. Su, H. Petersen, H. St\"{o}cker, and X. N. Wang, Nat. Commun. \textbf{9}, 210 (2018).
\bibitem{Brehmer2018PRL}J. Brehmer, K. Cranmer, G. Louppe, and J. Pavez, Phys. Rev. Lett. \textbf{121}, 111801 (2018).
\bibitem{Carrasquilla2017NP}J. Carrasquilla and R. G. Melko, Nat. Phys. \textbf{13}, 431 (2017).
\bibitem{Carleo2017Science}G. Carleo and M. Troyer, Science \textbf{355}, 602 (2017).
\bibitem{Navarro2021ApJ}F. Villaescusa-Navarro, D. Angl\'{e}s-Alc\'{a}zar, S. Genel, D. N. Spergel, R. S. Somerville, R. Dave, A. Pillepich, L. Hernquist, D. Nelson, Paul Torrey \emph{et al}., Astrophys. J. \textbf{915}, 71 (2021).
\bibitem{Navarro2022ApJS}F. Villaescusa-Navarro, S. Genel, D. Angl\'{e}s-Alc\'{a}zar, L. Thiele, R. Dave, D. Narayanan, A. Nicola, Y. Li, P. Villanueva-Domingo, B. Wandelt \emph{et al}., Astrophys. J. Suppl. Ser. \textbf{259}, 61 (2022).
\bibitem{Bedaque2021EPJA} P. Bedaque, A. Boehnlein, M. Cromaz, M. Diefenthaler, L. Elouadrhiri, T. Horn, M. Kuchera, D. Lawrence, D. Lee, S. Lidia \emph{et al}., Eur. Phys. J. A \textbf{57}, 100 (2021).
\bibitem{Boehnlein2022RMP} A. Boehnlein, M. Diefenthaler, N. Sato, M. Schram, V. Ziegler \emph{et al}., Rev. Mod. Phys. \textbf{94}, 031003 (2022).
\bibitem{He2023SCPMA} W. B. He, Q. F. Li, Y. G. Ma, Z. M. Niu, J. C. Pei, and Y. X. Zhang, Sci. China Phys. Mech. Astron. \textbf{66}, 282001 (2023).
\bibitem{Pederson2022NRP} R. Pederson, B. Kalita, and K. Burke, Nat. Rev. Phys. \textbf{4}, 357 (2022).
\bibitem{Brockherde2017NC} F. Brockherde, L. Vogt, L. Li, M. E. Tuckerman, K. Burke, and K.-R. M\"{u}ller, Nat. Commun. \textbf{8}, 872 (2017).
\bibitem{Li2021PRL} L. Li, S. Hoyer, R. Pederson, R. Sun, E. D. Cubuk, P. Riley, and K. Burke, Phys. Rev. Lett. \textbf{126}, 036401 (2021).
\bibitem{Kirkpatrick2021Science} J. Kirkpatrick, B. McMorrow, D. H. P. Turban, A. L. Gaunt, J. S. Spencer, A. G. D. G. Matthews, A. Obika, L. Thiry, M. Fortunato, D. Pfau \emph{et al}., Science \textbf{374}, 1385 (2021).
\bibitem{Nagai2020npjCM} R. Nagai, R. Akashi, and O. Sugino, npj Comput. Mater. \textbf{6}, 43 (2020).
\bibitem{Nikolaus1992PRC} B. A. Nikolaus, T. Hoch, and D. G. Madland, Phys. Rev. C \textbf{46}, 1757 (1992).
\bibitem{Burvenich2002PRC}T. B\"{u}rvenich, D. G. Madland, J. A. Maruhn, and P.-G. Reinhard, Phys. Rev. C \textbf{65}, 044308 (2002).
\bibitem{Zhao2010PRC} P. W. Zhao, Z. P. Li, J. M. Yao, and J. Meng, Phys. Rev. C \textbf{82}, 054319 (2010).
\bibitem{Zhao2022PRC} Q. Zhao, Z. X. Ren, P. W. Zhao, and J. Meng, Phys. Rev. C \textbf{106}, 034315 (2022).
\bibitem{Liu2023PLB} Z. X. Liu, Y. H. Lam, N. Lu, and P. Ring, Phys. Lett. B \textbf{842}, 137946 (2023).
\bibitem{Zhao2012PRC} P. W. Zhao, L. S. Song, B. H. Sun, H. Geissel, and J. Meng, Phys. Rev. C \textbf{86}, 064324 (2012).
\bibitem{Zhang2014FP} Q. S. Zhang, Z. M. Niu, Z. P. Li, J. M. Yao, and J. Meng, Front. Phys. \textbf{9}, 529 (2014).
\bibitem{Lu2015PRC} K. Q. Lu, Z. X. Li, Z. P. Li, J. M. Yao, and J. Meng, Phys. Rev. C \textbf{91}, 027304 (2015).
\bibitem{Zhang2021PRC} K. Y. Zhang, X. T. He, J. Meng, C. Pan, C. W. Shen, and S. Q. Zhang, Phys. Rev. C \textbf{104}, L021301 (2021).
\bibitem{SuppleMater} See Supplemental Material at \url{https://xxx} for detailed information on the CDFT with a point-coupling interaction and the neural network architecture.
\bibitem{Niu2018PLB} Z. M. Niu and H. Z. Liang, Phys. Lett. B \textbf{778}, 48 (2018).
\bibitem{Adam2014arXiv} D. P. Kingma and J. L. Ba, arXiv:1412.6980 [nucl-th].
\bibitem{PyTorch} PyTorch documentation, \url{https://pytorch.org/docs/stable/index.html}.
\bibitem{Wang2021CPC} M. Wang, W. J. Huang, F. G. Kondev, G. Audi, and S. Naimi, Chin. Phys. C \textbf{45}, 030003 (2021).
\bibitem{Utama2016PRC} R. Utama, J. Piekarewicz, and H. B. Prosper, Phys. Rev. C \textbf{93}, 014311 (2016).
\bibitem{Niu2022PRC} Z. M. Niu and H. Z. Liang, Phys. Rev. C \textbf{106}, L021303 (2022).
\bibitem{Wu2021PLB} X. H. Wu, L. H. Guo, and P. W. Zhao, Phys. Lett. B \textbf{819}, 136387 (2021).
\bibitem{Guo2024PRC} Y. Y. Guo, T. Yu, X. H. Wu, C. Pan, and K. Y. Zhang, Phys. Rev. C \textbf{110}, 064310 (2024).
\bibitem{Phillips2021JPG} D. R. Phillips, R. J. Furnstahl, U. Heinz, T. Maiti, W. Nazarewicz, F. M. Nunes, M. Plumlee, M. T. Pratola, S. Pratt, F. G. Viens, and S. M. Wild, J. Phys. G: Nucl. Part. Phys. \textbf{48}, 072001 (2021).
\bibitem{Qiu2024PLB} M. Qiu, B.-J. Cai, L.-W. Chen, C.-X. Yuan, and Z. Zhang, Phys. Lett. B \textbf{849}, 138435 (2024).
\bibitem{Kejzlar2020JPG} V. Kejzlar, L. Neufcourt, W. Nazarewicz, and P.-G. Reinhard, J. Phys. G: Nucl. Part. Phys. \textbf{47}, 094001 (2020).
\bibitem{Neufcourt2020PRC} L. Neufcourt, Y. Cao, S. Giuliani, W. Nazarewicz, E. Olsen, and Oleg B. Tarasov, Phys. Rev. C \textbf{101}, 014319 (2020).
\bibitem{Sharma2025PRR} A. Sharma, N. Schunck, and K. Wendt, Phys. Rev. Research \textbf{7}, 023179 (2025).

\end{thebibliography}
\end{document}